# Direct probing of phonon mode specific electron–phonon scatterings in two-dimensional semiconductor transition metal dichalcogenides: Symmetry and Berry phase

Duk Hyun Lee[1§], Sang-Jun Choi[2§], Hakseong Kim[1], Yong-Sung Kim[1] and Suyong Jung[1*]

[1]Korea Research Institute of Standards and Science, Daejeon 34113, Korea
[2]Institute for Theoretical Physics and Astrophysics, University of Würzburg, D-97074 Würzburg, Germany

## Abstract

Electron–phonon scatterings in solid-state systems are pivotal processes in determining many key physical quantities such as charge carrier mobilities and thermal conductivities. Here, we report on the direct probing of phonon mode specific electron–phonon scatterings in layered semiconducting transition metal dichalcogenides $WSe_2$, $MoSe_2$, $WS_2$, and $MoS_2$ through inelastic electron tunneling spectroscopy measurements, quantum transport simulations, and density functional calculation. We experimentally and theoretically characterize momentum-conserving single- and two-phonon electron–phonon scatterings involving up to as many as eight individual phonon modes in mono- and bilayer films, among which transverse, longitudinal acoustic and optical, and flexural optical phonons play significant roles in quantum charge flows. Moreover, we observe that two-phonon inelastic electron tunneling processes, which are confirmed to be generic in all four semiconducting layers, are governed by layer-number dependent symmetry, quantum interference, and geometric Berry phase.

[§] These authors contributed equally to this work.

**Introduction**

The collective vibrational modes in atomically arranged structures, namely phonons, and their interactions with charged carriers play crucial roles in determining various properties of condensed matter systems, covering thermal capacity and conductivity, electron mobility, and superconductivity, to name a few.[1–3] Specifically, in two-dimensional (2D) van der Waals (vdW) layered materials, which have drawn significant research interests not only for low-dimensional electrical and optical device applications but also as playgrounds to explore the topological phases of materials with non-trivial Berry curvatures,[4–7] electron–phonon interactions have been widely recognized as key elements in characterizing electronic, optical, and quantum properties.[8,9] Among those, electron–phonon scatterings are considered to determine the intrinsic charge carrier mobility in 2D semiconducting transition metal dichalcogenides (SC-TMDs), which has led to several theoretical studies on electron–phonon interactions in these types of layered materials.[10–15] For instance, as schematically illustrated in Figure 1a, two phonon modes $E''$ and $A_2''$ are optically accessible but weakly interact with electrons. In comparison, $E'$ and $A_1$ phonons are Raman active while strongly interacting with electrons. In particular, transverse and longitudinal acoustic phonons have been regarded as the primary factors that limit in-plane charge carrier mobilities in 2D vdW SC-TMD films.[12,16] Direct experimental approaches to explore electron–phonon interactions, however, have been largely missing, and most previous reports on phonon-related phenomena have been limited to either one or two isolated phonon modes and their temporal interactions with optically pumped hot electrons, which would not be relevant in active electronic applications.[17,18]

Free from stringent optical selection rules, inelastic electron tunneling spectroscopy (IETS) is known for its effectiveness in detecting electron–phonon interactions with a high energy resolution.[19–23] Previously, scanning tunneling microscopy (STM) has been used for local IETS measurements of a few phonon excitations in 2D semimetals.[20,21,24,25] For instance, van Hove singularities of graphene phonon bands and phonon-mediated inelastic channels to graphene have been observed under an STM probe.[20,21] When it comes to insulating 2D SC-TMDs, however, IETS studies with local tunnel probes have been limited due to weak tunnel signals from the point-like metallic probe to the gapped SC-TMDs. Moreover, IETS measurements through the atomically sharp tunnel junctions suffer from the intrinsic limitation of broad momentum spectra, making it further daunting to isolate individual phonon mode specific electron–phonon scatterings from concerted scattering networks on the Fermi surface of the 2D materials.

In this article, we report IETS measurements with four prototypical type-VI 2D SC-TMD films as tunnel media and find out that momentum-conserving electron–phonon scattering processes in 2D semiconducting films are governed by layer-number dependent electronic structures, symmetry, quantum interference, and the quantal geometric phase, *i.e.*, Berry phase. We experimentally and theoretically characterize single- and two-phonon IETS features involving up to as many as eight distinctive phonon modes in mono- and bilayer films. We find out that transverse (TA) and longitudinal (LA) acoustic phonons excited at the high symmetry points of Q, M, and K strongly interact with charged carriers in the 2D semiconductors. Several transverse (TO), longitudinal (LO), and flexural (ZO) optical phonons are also found to induce strong electron–phonon coupling, with each mode confirmed to have layer-number variant coupling strengths. Moreover, we identify several two-phonon inelastic electron tunneling processes that are governed by layer-number dependent symmetry, quantum interference, and the Berry phase. We confirm that the two-phonon IETS signals are generic in the four notable 2D semiconducting films, and reveal that they are highly sensitive to quantum interference that can be either constructive or destructive to the quantum charge flows depending on the Berry phase. Our measurements, corroborated with quantum transport simulation and density functional perturbation theory (DFPT), experimentally verify that momentum conservations, quantum interference, and the symmetry of the phonon excitations are essential in electron–phonon scattering processes and thus charged carrier transport in the 2D SC-TMD films.

## Results

**Inelastic electron tunneling spectroscopy with 2D vdW planar tunnel junctions.** A simplified device scheme is illustrated in the inset of Figure 1g. We implement thin graphite films (> 5 nm) as source and drain electrodes to preserve the intrinsic physical and electrical properties of the mono- and bilayer SC-TMD films, while introducing vdW tunnel barriers at the 2D vertical heterojunctions.[6,26,27] Another advantage of utilizing graphite as contact electrodes is to minimize the momentum mismatch of the electrons tunneling to and from the 2H-SC-TMD films, which enables efficient monitoring of the electron–phonon scattering processes around the K (Figure 1b) and Q (Figure 1c) points, in which the conduction band edges of the hexagonal 2D semiconducting layers are located.[6] As schematically illustrated in Figure 1d, electrons injected through a barrier transiently excite to the conduction band edges of the tunneling insulator. Although a majority of the electrons elastically tunnel through the insulator while releasing the pre-borrowed excitation energy (dashed and solid red lines in Figure 1d), electron–phonon interactions allow some electrons to be scattered off to other electronic states with or without a momentum change (solid blue and green wavy arrows in Figure 1d, respectively). Accordingly, adjunct transport channels are constructively established in the charge flows through the tunnel junctions, through which inelastic quantum tunneling events are exhibited as conductance modulations ($G = dI/dV_b$, Figure 1e) or as peaks or dips of the second derivative of tunnel current (d$G$/d$V_b$, Figure 1f) depending on sample bias ($V_b$) polarities. During these inelastic electron tunneling processes, momentum conservations limit the accessible phonons to the excitations at specific high-symmetric points, so that phonon excitations probed by our inelastic quantum tunneling measurements can be directly associated with the elemental electron–phonon scattering processes at the high-symmetric points and essentially govern the charge flows through the tunnel media.

Figure 1g–i show a collection of inelastic electron tunneling spectra from the first set of mono- (solid blue lines) and bilayer (dotted red lines) WSe$_2$ planar tunnel junctions at $T$ = 5.7 K. We add an AC excitation voltage ($V_{pp}$ = 1 mV, $f$ = 43.33 Hz) to DC $V_b$ and simultaneously measure both $I$–$V_b$ (Figure 1g) and $G$ = d$I$/d$V_b$–$V_b$ (Figure 1h) with an AC lock-in amplifier.[23] Within the $V_b$ range of |$V_b$| ≤ 100 mV, vertical charge flows through our graphite–WSe$_2$–graphite planar junctions are governed by direct tunneling, which is sensitive to WSe$_2$ layer thickness, tunnel junction area, crystallographic misalignment angles between the graphite electrodes and WSe$_2$ tunnel insulator, and electron–phonon scatterings. As explained before, weak IETS signals in d$I$/d$V_b$ (Figure 1e and 1h) can be better displayed as peaks and valleys in the second derivative of the tunnel current, d$G$/d$V_b$ = d$^2I$/d$V_b^2$ (Figure 1f and 1i), with their symmetric locations around $V_b$ = 0 mV representing the characteristic IETS feature directly linked to electron scatterings with phonons of activation energy $eV_b$.[19] It is worth pointing out that the shapes of the d$G$/d$V_b$ spectra are sensitive to the tunnel-junction parameters, so we limit our discussions to the devices with weak couplings, free from strongly-coupled resonance tunnel features (Figure 1d and 1i).[28] We obtained the d$G$/d$V_b$ spectra by numerically differentiating $G$ = d$I$/d$V_b$, confirmed to be consistent with data independently measured from an additional AC lock-in amplifier synchronized at a frequency of 2$f$ (Supplementary Figure 1). The conspicuous d$G$/d$V_b$ signals in our WSe$_2$-based

planar junctions indicate that tunnel electrons heavily interact with WSe₂ phonons, while the differing $dG/dV_b$ values and $V_b$ positions between the mono- and bilayer devices imply that electron–phonon scatterings in the 2D vdW heterostructures are WSe₂ layer-number variant. We note that all IETS spectra presented in the manuscript are obtained by numerically averaging out as many as 121 individual $dG/dV_b$ spectra, often varying external gate voltage $V_g$ applied to the Si/SiO₂/h-BN back gate (Supplementary Figure 2). By following this procedure, we can identify $V_g$-invariant IETS signals as outstanding peaks or dips, while the $V_g$-dependent features, such as those relating to defect states and the electronic structures of the SC-TMDs and graphite electrodes, can be avoided.[23]

**Single-phonon electron–phonon scatterings in mono- and bilayer WSe₂.** Figure 2b and 2c respectively show comprehensive IETS spectra from the second set of mono- and bilayer WSe₂ tunnel devices, measured at $T = 0.45$ K with an excitation voltage of $V_{pp} = 0.3$ mV. In total, we are able to identify eight independent IETS features in both mono- and bilayer samples within an energy window of $|eV_b| \leq 38$ meV and then compare them with DFPT-calculated phonon dispersions of freestanding monolayer WSe₂ (Figure 2a) and graphene–WSe₂–graphene heterostructures (Supplementary Figure 3). The $V_b$ positions for each IETS feature are assessed through a multi-peak Lorentzian fitting (solid grey curves in Figure 2b), with the collective fitting result (overlaid orange line in Figure 2b) excellently matching the experimental data. Moreover, the open blue squares in Figure 2b and 2c respectively indicate the experimental data in negative $V_b$ from the mono- and bilayer devices, with the $dG/dV_b$ dip locations perfectly aligning with the $dG/dV_b$ peaks in positive $V_b$. The widths of the ticks (Figure 2b and 2c) indicate the intrinsic IETS spectra-broadening in our measurements: FWHM ≈ 0.6 meV.[29]

Now we undertake the identification of the phonon modes responsible for each IETS d$G$/d$V_b$ peak in the monolayer WSe$_2$ device. Since momentum-conserving virtual quantum tunneling primarily occurs at the K and Q valleys at low temperatures, high-symmetric phonon modes connecting the K and Q valleys are expected to contribute strongly in the electron–phonon scattering processes in SC-TMDs. In that regard, our IETS measurements cannot be directly associated with the phonon density of states of thin films, for which phonon momentum is no longer a well-defined physical quantity. Recent theoretical studies also suggest that electron–phonon coupling strength, which is the crucial factor in determining IETS signals, is particularly strong at the high-symmetric points.[10–15] First, we mark the d$G$/d$V_b$ peak labeled as P$_2$ (8.57 mV ± 0.18 mV) in Figure 2b as associated with the TA-phonon branch, or more specifically, TA(Q) phonon excitation, which is the high-symmetry phonon mode exclusively excited within our measurement uncertainty. Similarly, P$_3$ (13.71 mV ± 0.07 mV) in the monolayer junction can be assigned to the LA(Q) phonons, which are predicted to cause strong electron–phonon interactions.[10–15] As a reference, transverse acoustic (ZA) modes are not expected to strongly deform electrostatic potentials, making electron scatterings with ZA phonons irrelevant in our IETS measurements. The next signal, P$_4$ (15.95 mV ± 0.05 mV), is within the phonon excitation energies of LA(K) and LA(M). Here, it is worth mentioning that the IETS features marked with P$_1$ in the monolayer (2.36 mV ± 0.12 mV) and bilayer (3.80 mV ± 0.13 mV) are associated with a newly excited lattice vibration mode in the graphite–WSe$_2$–graphite heterojunctions. Such low-energy excitations can be related to the intricate crystallographic arrangements of 2D vdW interfaces.[23]

Among the other six optical phonon branches of WSe$_2$ monolayers, LO$_2$, TO$_2$, and ZO$_1$ phonon modes are theoretically expected to generate strong electron–phonon scatterings.[10–15]

Accordingly, with relative ease, we can assign P$_5$ (24.63 mV ± 0.08 mV) in the monolayer to the LO$_2$(K) phonon mode. Identifying the primary phonon excitations for the next three d$G$/d$V_b$ peaks, however, is not as straightforward as those for the previously assigned low-energy phonons. For example, P$_6$ (27.48 mV ± 0.04 mV) is in close proximity to LO$_2$(M)/TO$_2$(M) and LO$_2$(Q), while P$_7$ (31.33 mV ± 0.07 mV) and P$_8$ (33.45 mV ± 0.06 mV) are within the energy ranges of the TO$_2$ (Γ, K)/ZO$_1$(Γ, K) and ZO$_1$(Q, M) phonon excitations. Here, we point out that identifying the primary optical modes from these energetically close-packed WSe$_2$ optical phonon excitations can be possible with two-phonon scattering IETS measurements and quantum transport simulations, as we discuss in detail later.

We find that the IETS spectra from the bilayer device, thus the electron–phonon scattering characteristics, are distinct when compared with their monolayer counterparts. Most of all, inversion symmetry preserved in bilayer 2H SC-TMDs renders interlayer tunneling at the K (K′) valley irrelevant since the interlayer couplings at the K (K′) points are weak around the conduction band edges where the electronic bands of SC-TMDs are mostly composed of d$_z^2$ orbitals.[7,30,31] In comparison, strong interlayer hybridizations are prompted by the orbitals responsible for the conduction band edges at Q. Thus, electrons tunneled into the first layer of the bilayer from the graphite electrode with momentum K (K′) should be scattered off to Q (Q′) through intra- or intervalley electron–phonon scatterings (Figure 1b) before tunneling to the other layer.[7,30] Accordingly, single-phonon interlayer tunneling assisted by K phonons becomes limited in SC-TMD bilayers, leading to diminished inelastic tunnel features with the phonons at K. As marked with red inverted triangles in Figure 2c, the electron–phonon scattering signal with LO$_2$(K) phonons is indeed reduced in intensity in the bilayer (P$_5$ in Figure 2c) when compared with the conspicuous IETS feature for LO$_2$(K) phonons in the monolayer (Figure 2b). Similarly, the d$G$/d$V_b$ hump marked as P$_4$ (15.60 mV ± 0.28 mV) becomes attenuated in intensity, and the moderate undertone of the d$G$/d$V_b$ spectra, P$_7$ in the bilayer, could be attributed to quenched TO$_2$(K) and ZO$_1$(K) phonon excitations as well.

Thanks to an excellent tunnel-junction stability, moreover, we are able to probe high-energy inelastic electron tunneling processes in 2D vdW SC-TMDs. Figure 3a and 3b respectively show d$G$/d$V_b$ spectra from the mono- and bilayer WSe$_2$ devices up to |$eV_b$| ≤ 70 meV. Note that no available WSe$_2$ phonon modes exist within the energy range 40 meV ≤ |$eV_b$| ≤ 70 meV, with the majority of phonons associated with the graphite electrodes having much higher energies.[23,32] Interestingly, these high-energy IETS spectra differ by $eV_b$ location between the mono- and bilayer WSe$_2$ devices. In the monolayer, for example, a distinct d$G$/d$V_b$ peak is observed at ≈ 59 meV along with a rather broad d$G$/d$V_b$ hump at ≈ 53 meV. Meanwhile, as shown in Figure 3b for the bilayer, the strongest IETS signal is formed at ≈ 42 meV, at which no apparent d$G$/d$V_b$ spectra exist in the monolayer WSe$_2$. In addition, the high-energy spectra in the bilayer (40 meV ≤ |$eV_b$| ≤ 60 meV) are far stronger in intensity than the IETS signals at low energies (|$eV_b$| ≤ 40 meV), which is not the case in the monolayer device. Here, we accredit the sets of higher-energy layer-number-dependent IETS spectra to two-phonon electron–phonon scatterings that are governed by quantum interference and the Berry phase, which we elaborate on in the next section.

**Quantum interference and Berry phase in two-phonon electron–phonon scattering processes in SC-TMDs.** We find out that quantum interference and the Berry phase play crucial roles in understanding the distinct higher-energy inelastic electron scattering processes in SC-TMDs. Let us begin with an intuitive description of quantum interference and the Berry phase in two-phonon electron–phonon scatterings, during which an electron with an initial momentum $k_i$ is interacting with two respective phonons of momenta $q$ and $q'$ and ends up in a state with a final momentum of $k_f = k_i - q - q'$. In a microscopic view, momentum conservation allows two different electron–phonon inelastic processes with differing scattering orders: emitting the first $q$ ($q'$) phonon and stopping at the intermediate state $\kappa_A = k_i - q$ ($\kappa_B = k_i - q'$), and arriving at $\kappa_f$ after scattering off by emitting the second $q'$ ($q$) phonon. Note that such a pair of distinctive scattering routes forms a closed loop in momentum space, and the developed quantum superposition finally determines the inelastic electron tunneling probability that is responsible for the experimentally observable IETS signals in magnitude (insets in Figure 3c and 3d).

Quite intriguingly, such a quantum interference effect comes into play for various two-phonon electron–phonon scattering processes, owing to the peculiar electronic structures of the SC-TMDs with six distinct Q and K valleys around the conduction band edges. For instance, when scattered by M and Q phonons such as LA(M) and LO(Q), an electron at the K point is allowed to travel through two intermediate states ($\kappa_A, \kappa_B$) at Q$_1$ and Q$_4$ before arriving at the K′ point (inset in Figure 3d). During these scattering processes, the quantum interference effect becomes prominent since the probability of each scattering route is the same thanks to the identical band structures around the Q$_1$ and Q$_4$ points. In stark contrast, however, when tunnel electrons are scattered by the phonons at M and K, during which the electrons detour through the intermediate states at the K and Q valleys, the quantum interference effect becomes attenuated due to the dissimilar tunneling probabilities resulting from the differing energy gaps at the K and Q valleys (inset in Figure 3c).

When a sudden spin change accompanies the course of electrons traveling through a closed loop in momentum space, moreover, the Berry phase becomes a key factor in determining quantum interference, as proven in interferometry measurements of neutron scattering, optics, and others.[33–37] Similarly, the unique spin-momentum locking and impactive momentum changes of the tunnel electrons here require us to pay close attention to the Berry phase in the course of the two-phonon inelastic electron–phonon scattering in the SC-TMDs. With spin-1/2 electrons, the Berry phase equals half of the solid angle of the geodesic polygon connecting the electronic spin states of $k_i$, $k_f$, $\kappa_A$, and $\kappa_B$ on the Bloch sphere. More importantly, the Berry phases, similarly to the electronic states in mono- and bilayer SC-TMD films, are determined by the symmetries present in 2D crystalline films. In the monolayer SC-TMDs where time-reversal symmetry is preserved but inversion symmetry is not, the $\pi$ Berry phase is added to the accumulated quantum interference in the two-phonon inelastic electron scattering processes with M and Q phonons. Since the spin states of the time-reversal counterparts of $k_i$ and $k_f$ ($\kappa_A$ and $\kappa_B$) are located at the antipodal points on the Bloch sphere, the solid angle of the geodesic polygon connecting the opposite spin states is $2\pi$, which leads to destructive quantum interference in the monolayer SC-TMDs with an additional $\pi$ Berry phase (Supplementary Note). On the other hand, the simultaneous presence of both time-reversal and inversion symmetries in the SC-TMD bilayers forces the Berry phase to vanish, and accordingly, the quantum phase around the closed loop is equivalently $2\pi$, resulting in constructive quantum interference in the bilayer films. In general, the Berry phase is calculated by integrating the Berry curvature $\Omega(\vec{k})$, and with the time-reversal and inversion symmetries giving $\Omega(\vec{k}) = -\Omega(-\vec{k}) = -\Omega(\vec{k})$, we get $\Omega(\vec{k}) = 0$.[38,39]

With a rigorous quantum mechanical IETS simulation (see Methods and the Supplementary Note for detailed descriptions), we verify that the higher-energy two-phonon IETS signals are indeed governed by the aforementioned quantum interference and Berry phase. We consider all possible combinations of two-phonon inelastic scattering routes out of the experimentally identified eight individual phonon modes, and verify that the two-phonon IETS signals specifically associated with Q and M phonons are highly sensitive to the layer-number dependent symmetries and quantum interference. We figure vertical charge flow through the bilayer as interlayer tunneling through two WSe$_2$ films with the exclusion of intervalley scattering around the K (K´) valley in the scattering matrix. Figure 3c and 3d display simulated d$G$/d$V_b$ spectra for mono- and bilayer WSe$_2$ with full consideration of the Berry phase and quantum interference around the closed loop, and the agreement with the experimental data is outstanding—in particular, the absence (presence) of d$G$/d$V_b$ signals within 40 meV $\leq |eV_b| \leq$ 55 meV in the mono- (bi-) layer WSe$_2$ films. We provide all the theoretically expected positions of IETS features and two-phonon inelastic scattering processes exhibiting quantum interference in Supplementary Table 1. It should be pointed out that our simplified transport model for bilayer WSe$_2$ has shortcomings to fully analyze the experimental observations. For example, the two-phonon inelastic electron scatterings with LA phonons are expected to significantly contribute to the vertical charge flows, resulting in higher d$G$/d$V_b$ spectra in value around 25 meV $\leq |eV_b| \leq$ 35 meV (dotted orange line in Figure 3d). Instead, the simulated spectra without considering the double LA phonon scatterings (solid blue line in Figure 3d) are found to be close to the experimental observations, calling for further theoretical works to better clarify the phonon interactions with conducting electrons in bilayer SC-TMDs.

**Electron-phonon scatterings in mono- and bilayer MoS$_2$, MoSe$_2$ and WS$_2$.** We find out that the key electron–phonon scattering characteristics observed with WSe$_2$ films—that single- and two-phonon electron–phonon scatterings are regulated by layer-number dependent symmetries and Berry phases—are generic to the other type-VI SC-TMD films, namely $MoS_2$, $MoSe_2$, and $WS_2$. Figure 4b shows an IETS spectrum from a monolayer $MoS_2$ tunnel device, measured at $T = 0.45$ K with an excitation voltage of $V_{pp} = 0.5$ mV. From the monolayer $MoS_2$, we can identify six distinct IETS features within an energy window of $|eV_b| \leq 55$ meV, with each spectrum closely aligned to the TA, LA, $LO_2$, $TO_2$, and $ZO_1$ phonon branches (Figure 4a). Identical to the monolayer $WSe_2$ films, TA(Q), LA(Q), and LA(M, K) phonons can be marked as the leading acoustic phonon excitations that generate sizable electron–phonon scatterings in monolayer $MoS_2$ films. Separated by a sizable phonon gap at 30 meV $\leq |eV_b| \leq$ 45 meV, three $dG/dV_b$ peaks appear close together and can be linked to $LO_2$, $TO_2$, and $ZO_1$ optical phonon branches. Moreover, similar to the high-energy two-phonon IETS features in the monolayer $WSe_2$, there exist several weak but distinct $dG/dV_b$ peaks within the energy range, 50 meV $\leq |eV_b| \leq$ 100 meV. As discussed previously, the IETS spectra, thus electron–phonon scatterings, in bilayer $MoS_2$ films contrast those in monolayer $MoS_2$. Figure 4c shows an IETS spectrum from a bilayer $MoS_2$ tunnel device, and the high-energy IETS features at 55 meV $\leq |eV_b| \leq$ 100 meV, which are attributed to the two-phonon inelastic scatterings, are higher in intensity than those for the low-energy IETS signals. In addition, as similar to bilayer $WSe_2$, the IETS feature corresponding to the single-phonon electron scatterings with LA(K) phonons becomes diminished in the bilayer $MoS_2$, as marked with an inverted red triangle in Figure 4c.

The vertical red tick marks and colored solid and dotted arrows in Figure 4c indicate various two-phonon inelastic electrons scattering processes in the bilayer, during which quantum phases are accumulated constructively, forming well-defined d$G$/d$V_b$ peaks at assorted MoS$_2$ two-phonon energies with Q and M phonons. In comparison, as marked with vertical blue tick marks in Figure 4b for the monolayer, most high-energy d$G$/d$V_b$ peaks can be explained by the two-phonon electron–phonon scatterings, save for the destructive Q and M combinations. Based on these observations, we can infer that the elementary phonon modes that limit the intrinsic charge carrier mobility in MoS$_2$-based electronic devices are TA(Q), LA(Q), LA(M), LO$_2$(Q)/LO$_2$(M), TO$_2$(Q)/TO$_2$(M), and ZO$_1$(M) phonons.

Figure 5c and 5e respectively display IETS measurements from the mono- and bilayer MoSe$_2$ tunnel devices, and Figure 5d and 5f are those from the WS$_2$ mono- and bilayer devices, along with the DFPT-calculated phonon dispersions and phonon density of states of freestanding monolayer MoSe$_2$ (Figure 5a) and WS$_2$ (Figure 5b). All IETS spectra are measured at $T = 0.45$ K with an excitation voltage of $V_{pp} = 0.5$ mV. As displayed in our measurements, the tunnel spectra observed in MoSe$_2$ and WS$_2$ are consistent with the previously discussed electron–phonon scattering physics in WSe$_2$ and MoS$_2$, in particular, inversion symmetry-regulated charged carrier scattering with K phonons: LA(K) as indicated with inverted red triangles in Figure 5e and 5f, and Berry phase administered Q and M two-phonon inelastic scatterings. Some of the prominent two-phonon modes for MoSe$_2$ and WS$_2$ are marked with colored arrows and their combinations in Figure 5c–f, with the primary single-phonon excitations identified in our measurements denoted with colored stars in Figure 5a for MoSe$_2$ and Figure 5b for WS$_2$. It is interesting to point out that, unlike the two-phonon excitation modes in WSe$_2$, MoS$_2$, and MoSe$_2$, the most prominent d$G$/d$V_b$ peak in bilayer WS$_2$ is located at ≈ 32 meV because a sizable phonon gap is formed between the acoustic and optical phonon branches; we can attribute such a strong peak to two-phonon electron scatterings with acoustic phonons of TA(Q) + LA(M) and TA(M) + LA(Q). Lastly, we remark that the experimentally identified $ZO_1$ phonons are consistently higher in energy than the theoretically expected freestanding SC-TMD phonon excitations, as indicated with red arrows in Figure 2a ($WSe_2$), Figure 4a ($MoS_2$), Figure 5a ($MoSe_2$), and Figure 5b ($WS_2$), suggesting that flexural motions of chalcogen atoms become hardened in 2D vdW vertical heterostructures by as much as $\approx 3$ meV.[23] Although it is not straightforward to draw the exact phonon dispersions of our graphene–SC-TMD–graphene heterojunctions, primarily due to the lattice mismatches, we discover that the assessable ZO phonon density of states indeed shifts to higher energies, even in the simplest graphene(4×4)–$WSe_2$(3×3) heterostructures (Supplementary Figure 3). In total, we measured four mono- and four bilayer $WSe_2$ devices, with all showing consistent two-phonon electron–phonon scattering features with Q and M phonons (Supplementary Figures 4 and 5). Our findings were additionally confirmed with three mono- and two bilayer $MoSe_2$, two mono- and three bilayer $WS_2$, and one mono- and one bilayer $MoS_2$ planar tunnel junctions. IETS spectra for the devices not discussed in the main text are presented in Supplementary Figures 6 and 7.

**Discussions**

To further support our findings on momentum-conserving inelastic electron–phonon scattering processes in SC-TMDs, we prepared another type of graphite–SC-TMD–graphite vertical heterostructure: twisted double-layer $WSe_2$ vertical junctions. As shown in the optical image in Supplementary Figure 8, two monolayer $WSe_2$ films are serially transferred on top of the bottom graphite flake, and the misalignment angle of the double $WSe_2$ layers is estimated to be around 13° as judged from the crystallographic directions of each layer. Distinct from Bernal stacked SC-TMD bilayers, the inversion symmetry of the twisted double-layer $WSe_2$ film is naturally broken such that the IETS features from the twisted double-layer device should be quite dissimilar to those from conventional bilayer WSe$_2$. Indeed, we find that the overall IETS signals in the inversion-symmetry-broken twisted WSe$_2$ double layers are similar to those from the monolayer devices in terms of the excited phonon modes, without noticeable d$G$/d$V_b$ features that may relate to the Q + M two-phonon excitations (Supplementary Figure 8). Although more in-depth experimental works should follow to clarify the compelling electron–phonon scatterings in twisted double-layer systems, we feel confident that the current data sufficiently supports our claims made in the current manuscript: inversion symmetry, quantum interference, and the Berry phase play important roles in both single- and two-phonon electron–phonon scattering processes in SC-TMD films.

We implement graphite flakes as the source and drain electrodes in our vertical planar tunnel junctions to preserve the intrinsic electronic properties of the SC-TMD layers. As electrons tunnel through the graphite–SC-TMD vdW vertical junctions, there is a high chance they will be scattered by graphite phonons as well. From our IETS measurements, we are indeed able to locate several d$G$/d$V_b$ spectra that are likely related to the graphite phonons and the two-phonon inelastic electron scatterings with the phonons of the SC-TMDs and graphite layers. For example, the d$G$/d$V_b$ spectra at ≈ 66 meV in both mono- and bilayer devices, as respectively marked with red arrows and dotted yellow lines in Figure 6a and 6b, are from the graphite ZO(K) phonons.[20,22,23] It is worth mentioning that the graphite ZO(K) signals persist up to $T \geq 100$ K, in stark contrast to the temperature sensitive two-phonon IETS features discussed above. Moreover, the graphite phonon modes emerge only when the two-phonon WSe$_2$ electron–phonon scatterings become diminished at elevated temperatures ($T > 30$ K). We are also able to observe several other graphite and two-phonon graphite–SC-TMD phonon signatures within the energy range $|eV_b| \leq 300$ meV, but detailed analyses of these high-energy spectra are beyond the current study and will be presented elsewhere. Finally, moving beyond bilayer, we find that the IETS signals originating from two-phonon-mode electron–phonon scatterings in three- and four-layer SC-TMD devices are consistently higher in intensity, as presented in Supplementary Figure 9, suggesting that charge flows through multi-layered SC-TMDs become heavily regulated and are often facilitated by multiple electron–phonon scattering processes.

In summary, we spectroscopically characterized phonon mode specific electron–phonon scatterings in four prototypical 2D semiconducting films, $WSe_2$, $MoS_2$, $WS_2$, and $MoSe_2$, by inelastic electron tunneling spectroscopy measurements, quantum transport simulations, and density functional perturbation theory. Thanks to the superb physical and electrical stability of our planar tunnel junctions, we were able to probe several single- and two-phonon inelastic electron scattering processes that are governed by layer-number dependent electronic structures, inversion symmetry, and Berry phase in the SC-TMD films. From the standpoint of novel electron–phonon scattering physics in condensed matter systems, our measurements are the first experimental demonstrations that quantum interference is a major player in momentum-conserving inelastic electron–phonon scatterings and thus charge transport behaviors in 2D SC-TMDs. In addition, we demonstrated that our experimental approach, utilizing inelastic tunneling spectroscopy with 2D planar vdW tunnel junctions as a high-fidelity material metrology platform, is applicable to a wide range of low-dimensional quantum materials and their unlimited combinations for probing charge carrier interactions with phonons and other intriguing quasiparticles.[40,41]

## Methods

**Device Fabrication.** In our planar vdW heterostructures, preparation of atomically clean interfaces is of critical importance for an accurate and reliable material characterization of the vertical junctions. At first, 60 nm to 100 nm thick *h*-BN flakes are mechanically exfoliated on a 90 nm thick SiO$_2$ layer on Si substrate. Then, a mechanically isolated graphite flake of thickness 5 nm or more is transferred to a pre-located *h*-BN flake on the SiO$_2$/Si substrate using a dry transfer method. We utilize polymer stacks of PMMA (poly(methyl methacrylate))–PSS (polystyrene sulfonate) layers for such tasks and carefully adjust the thickness of each layer to enhance the optical contrast of the exfoliated ultrathin 2D layered materials. We remove the PMMA film in warm (60 ºC) acetone and further anneal the samples at 350 ºC for several hours in a mixture of Ar : H$_2$ = 9 :1 to ensure residue-free graphite surfaces. Next, instead of again using the polymer stacks, we use a Gel-Pak to exfoliate and transfer mono- and bilayer SC-TMD films on top of the *h*-BN–graphite stack. Gel-Pak residue free surfaces are confirmed with an atomic force microscope measurement. Finally, a top graphite flake prepared on the PMMA–PSS polymer stack is transferred to form a vertical graphite–SC-TMD–graphite planar tunnel device. We note that the crystallographic angles of the graphite electrodes and the SC-TMD films are intentionally misaligned, and the thicknesses of the semiconducting layers are confirmed via atomic force microscope. The active junction areas, which are determined by the widths of the top and bottom graphite flakes, are several tenths of a square micrometer. We purchased high-purity (> 99.995%) SC-TMD crystals from HQ Graphene with no additional dopants added during growth procedures and large size graphenium flakes from NGS Naturgraphit GmbH.

**DFPT for calculating phonon dispersion.** Phonon dispersions of free-standing SC-TMD monolayers and graphene–WSe$_2$–graphene heterostructures are calculated using density functional perturbation theory (DFPT), implemented in Vienna Ab initio Simulation Package[42] within generalized gradient approximation (PBE).[43] Projector augmented pseudopotentials are used and the plane-wave cutoff is set to be 500 eV.[44] The phonon dispersions of free-stranding SC-TMD monolayers are calculated with a 3×3 supercell and 2×2 k-point mesh. The phonon structures of graphene–WSe$_2$–graphene heterostructures are calculated with a 3×3 WSe$_2$ supercell and 4×4 graphene supercell with 2×2 k-point mesh as well. The lattice constant of WSe$_2$ is set to be 3.325 Å in PBE, and the graphene lattices are relaxed by –1% to compensate for any WSe$_2$–graphene lattice mismatch.

**Quantum transport simulation.** Quantum transport simulations are performed using an electron-tunneling scattering matrix with two-particle Green functions. The two-particle Green function $G_{\kappa\kappa'}(\tau, s, t)$ determines the transmission probability of the tunnel junctions in the time domain as follows

$$T(\epsilon_f, \epsilon_i) = \sum_{\kappa,\kappa'} \int \int \int d\tau \, ds \, dt \, e^{\frac{i[(\epsilon_i-\epsilon_f)\tau+\epsilon_f t-\epsilon_i s]}{\hbar}} G_{\kappa\kappa'}(\tau, s, t),$$

where $\tau, s, t > 0$. The transmission probability $T(\epsilon_f, \epsilon_i)$ is used to calculate the vertical electron tunneling current, $I(V) \propto \int d\epsilon_f d\epsilon_i T(\epsilon_f, \epsilon_i)[f_L(\epsilon_i - eV) - f_R(\epsilon_f)]$. The Green function consists of the product of a probability amplitude and its complex conjugation, which correspond to the propagation of electrons moving forward and backward in the time domain. Although algebraic evaluation of the two-particle Green function is extremely complicated, Feynman diagrams as provided in Figure M1 simplify our calculations while providing intuitive understanding. Each vertex represents electron–phonon interaction with coupling strength $M^\lambda_{\kappa',\kappa}$. Here, $\lambda$ denotes a phonon mode that scatters an electron from momentum state $\kappa$ to $\kappa'$. When electrons scatter with the phonon modes $\lambda_r$ and $\lambda_b$, the momentum-conserving two-phonon inelastic electron scatterings allow four independent scattering processes, as depicted in the Feynman diagram (Figure M1).

Then, the transmission probability $T$ can be estimated with the absolute square of the sum of the two electron–phonon scattering amplitudes, *i.e.*, $T=|A+B|^2=|A|^2+|B|^2+A(B)^*+(A)^*B$, where the first and second terms represent the first and second Feynman diagrams in Figure M1. The third and fourth terms, which are responsible for the quantum interference, represent the third and fourth Feynman diagrams, respectively.

Now we discuss how the Berry phase appears in the quantal phase of the third and fourth terms in the transmission probability equation. When we evaluate the two-particle Green function to simulate vertical tunneling, electron–phonon scatterings for each scattering route are accompanied by the coupling strength $M^\lambda_{\kappa',\kappa}$, which can be expressed as

$$M^\lambda_{\kappa+q,\kappa} = \eta^\dagger_{\kappa+q}\eta_\kappa \int d\mathbf{r}\phi^*_{\kappa+q}(\mathbf{r})\delta V_{q\lambda}(\mathbf{r})\phi_\kappa(\mathbf{r}),$$

where $\eta_\kappa$ is the spinor of the Bloch wave function $\phi_\kappa(\mathbf{r})$, and $\delta V_{q\lambda}$ is the derivative of the effective potential with respect to the displacement induced by the phonon mode $\lambda$. Note that the pre-factor $\eta^\dagger_{\kappa+q}\eta_\kappa$ appears due to the spin-momentum locking in SC-TMDs, and its phase is obtained from the line integral of the Berry connection along the geodesic $C$ connecting the spin states of $\eta_{\kappa+q}$ and $\eta_\kappa$ on the Bloch sphere, *i.e.*, $\arg(\eta^\dagger_{\kappa+q}\eta_\kappa) = \int_C d\mathbf{R} \cdot \eta^\dagger(\mathbf{R})\nabla_R\eta(\mathbf{R})$. Once the second-order electron–phonon scattering processes form a closed loop in the momentum space, the sum of the relevant phase factors becomes the gauge-independent Pancharatnam–Berry phase that is generalized to discontinuous or noncyclic cases[35,45–47], so that the third and fourth terms of the transmission probability equation entail an additional phase factor form, $\left(M^{\lambda_b}_{\kappa',\kappa_B} M^{\lambda_r}_{\kappa_B,\kappa}\right)^* M^{\lambda_r}_{\kappa',\kappa_A} M^{\lambda_b}_{\kappa_A,\kappa} \propto (\eta^\dagger_\kappa\eta_{\kappa_B})(\eta^\dagger_{\kappa_B}\eta_{\kappa'})(\eta^\dagger_{\kappa'}\eta_{\kappa_A})(\eta^\dagger_{\kappa_A}\eta_\kappa)$. Especially, when $\eta_\kappa$ ($\eta_{\kappa_A}$) and $\eta_{\kappa'}$ ($\eta_{\kappa_B}$) are time-reversal partners, the quantal phase of $(\eta^\dagger_\kappa\eta_{\kappa_B})(\eta^\dagger_{\kappa_B}\eta_{\kappa'})(\eta^\dagger_{\kappa'}\eta_{\kappa_A})(\eta^\dagger_{\kappa_A}\eta_\kappa)$

is $\pi$. Detailed evaluations of $G_{\kappa\kappa'}(\tau, s, t)$ and the Hamiltonian are presented in the Supplementary Information.

**Data availability**

All data supporting the findings of this study are available from the corresponding authors on request.

**Acknowledgments.** This work was supported by research grants for basic research (KRISS-2020-GP20011059) funded by the Korea Research Institute of Standards and Science and the Basic Science Research Program (MRF-2019R1A2C2004007) through the National Research Foundation of Korea. This work was also supported by the DFG (SFB1170 "ToCoTronics"), the Wűzburg-Dresden Cluster of Excellence ct.qmat, EXC2147, project-id 39085490.

**Author contributions.** S.J. and S-J. C. designed the experiments, and D.H.L and H.K. fabricated the devices and performed the inelastic electron tunneling spectroscopy measurements. S-J. C. carried out the quantum transport simulations and Y-S. K. performed the DFPT calculations. D.H.L., S-J. C., Y-S. K., and S.J. analyzed the data and co-wrote the paper. All authors contributed to the manuscript.

**Competing interests.** The authors declare no competing interests.

**Additional information**

**Correspondence and requests for materials** should be addressed to S.J.

**Figure Captions**

**Figure 1. Inelastic electron tunneling spectroscopy probing electron–phonon scatterings in SC-TMD films. a**, Schematic viewgraphs of lattice vibration modes that are either weakly (upper) or strongly (lower) coupled to electrons in 2H-SC-TMD layers. **b**, **c**, Schematic illustrations of

electron–phonon scatterings initiated at either the K (**b**) or Q (**c**) valley in the momentum space of a 2D hexagonal lattice. **d**, Schematic of the energy-band alignment and quantum tunneling events in a vertical graphite–SC-TMD–graphite planar tunnel junction. Injected electrons within an energy window of $eV_b$ either elastically tunnel (red arrows) through the barriers or interact with SC-TMD phonons without (blue arrows) or with (green arrows) a momentum transfer before exiting to a drain electrode. **e**, **f**, Simplified viewgraphs of the evolutions of differential conductance ($G = dI/dV_b$, **e**) and a second derivative of tunnel current ($dG/dV_b = d^2I/dV_b^2$, **f**) as a function of sample-bias voltage ($V_b$). Experimental signatures of the single- and two-phonon electron–phonon couplings are represented as a $dI/dV_b$ increase (**e**) and a $dG/dV_b$ peak or dip (**f**) at individual phonon energies $\hbar\omega_1$ and $\hbar\omega_2$ and combined phonon-mode energy $\hbar\omega_3 = \hbar\omega_1 + \hbar\omega_2$. **g–i**, Inelastic electron tunnel spectra from a mono- (solid blue lines) and bilayer (dotted red line) WSe$_2$ planar tunnel junction at $T = 5.7$ K. Tunnel electron characteristics are represented in $I - V_b$ (**g**), $dI/dV_b - V_b$ (**h**), and $dG/dV_b - V_b$ (**i**) curves, respectively. (inset, **g**) Schematic of our graphite–SC-TMD–graphite vertical planar junction and two-probe electric measurement setup.

**Figure 2. Single-phonon electron–phonon scatterings in mono- and bilayer WSe$_2$.** **a**, DFPT-calculated phonon dispersion and phonon density of states of freestanding monolayer WSe$_2$. The phonon branches generating strong electron–phonon couplings, namely TA, LA, LO$_2$, TO$_2$, and ZO$_1$, are indicated with bold lines, and the optically accessible phonon modes at Γ, namely E″, E′, A$_1$, and A$_2$″, are marked with orange dots. Colored stars mark the experimentally verified phonon excitations that strongly couple to electrons. **b**, **c**, $dG/dV_b$ evolutions of the positive (open brown circles) and negative (open blue squares) $V_b$ in mono- (**b**) and bilayer (**c**) WSe$_2$ tunnel devices. The IETS spectra were taken at $T = 0.45$ K with an excitation voltage of $V_{pp} = 0.3$ mV.

**Figure 3. Two-phonon electron–phonon scatterings in mono- and bilayer WSe$_2$. a**, **b**, Detailed d$G$/d$V_b$ evolutions of positive (open brown circles) and negative (open blue squares) $V_b$ in mono- (**a**) and bilayer (**b**) WSe$_2$ tunnel devices. The IETS spectra were taken at $T$ = 0.45 K with an excitation voltage of $V_{pp}$ = 0.3 mV. The lengths of the colored arrows, used for guiding the two-phonon excitations, represent the primary phonon energies in the monolayer WSe$_2$. **c**, **d**, Simulated d$G$/d$V_b$ spectra from quantum transport calculations for mono- (**c**) and bilayer (**d**) WSe$_2$ films. The single-phonon excitations are based on the monolayer data (Figure 2b), and the two-phonon electron–phonon scatterings are from the composite two-phonon scattering routes regulated by layer-number dependent electronic band structures, inversion symmetries, and Berry phases. (insets, **c**, **d**) Schematic illustrations of two-phonon electron–phonon scattering routes at the K valley interacting with M and K phonons (**c**), and M and Q phonons (**d**).

**Figure 4. Single- and two-phonon electron–phonon scatterings in mono- and bilayer MoS$_2$.** **a**, DFPT-calculated phonon dispersion and phonon density of states of freestanding monolayer MoS$_2$. The graphite phonon branches are illustrated with thin grey lines. The phonon branches confirmed to generate strong electron–phonon scatterings are indicated with bold lines. **b**, **c**, d$G$/d$V_b$ evolutions as a function of energy in mono- (**b**) and bilayer (**c**) MoS$_2$ tunnel devices. The IETS spectra were taken at $T$ = 0.45 K with an excitation voltage of $V_{pp}$ = 0.5 mV. The experimentally identified single phonon-mode energies are distinctly marked with solid and dotted colored arrows, and the two-phonon electron–phonon scattering features are indicated with vertical blue (**b**) and red (**c**) tick marks, guided with vertical grey lines.

**Figure 5. Electron–phonon scatterings in mono- and bilayer MoSe₂ and WS₂.** **a**, **b**, DFPT-calculated phonon dispersions and phonon density of states of freestanding monolayer MoSe₂ (**a**) and WS₂ (**b**). **c**–**f**, d$G$/d$V_b$ evolutions as a function of energy in mono- and bilayer MoSe₂ and WS₂ tunnel devices. The IETS spectra were taken at $T = 0.45$ K with an excitation voltage of $V_{pp} = 0.5$ mV. The experimentally identified single phonon mode energies are marked with solid and dotted colored arrows, and the two-phonon electron–phonon scatterings are indicated with vertical tick marks.

**Figure 6. Temperature-dependent IETS spectra in mono- and bilayer WSe₂.** **a**, **b**, Two-dimensional display of d$G$/d$V_b$ evolutions of the first set of mono- (**a**) and bilayer (**b**) WSe₂ tunnel devices as a function of $V_b$ at varying temperatures from $T = 6$ K to $T = 100$ K with $\Delta T = 2$ K. Dotted yellow and magenta arrows indicate the $T$-dependent evolutions of the WSe₂ two-phonon electron–phonon scatterings to the graphite ZO(K) phonons at $\approx 66$ meV.

**Methods Figure Caption**

**Figure M1. Feynman diagrams of $G^2_{\kappa\kappa'}(\tau, s, t)$ for the second–order peaks of d$G$/d$V_b$.** The solid lines with right and left arrows represent the propagations of electrons moving forward and backward in the time domain in the conduction band, respectively. The backward propagation appears as a complex conjugation of the probability amplitude in the forward propagation. Colored wavy lines denote the exchange of phonons between electrons and holes, while different colors refer to different phonon modes such as LA and LO. Two different tunneling processes via intermediate states of $\kappa_A$ and $\kappa_B$ exhibit quantum interference when those processes involve two

phonons with a reversed order. The interaction strength of each term is presented before the parentheses. We highlight that quantum interference occurs as the intermediate states $\kappa_A$ and $\kappa_B$ are distinct in 2D SC-TMDs.